\documentstyle[jprart,epsf,times,journals]{aipproc}

\begin{document}

\title{Hadronic Correlated Flares from Mrk 501}

\author{J\"org P. Rachen}

\address{Sterrenkundig Instituut, Universiteit Utrecht, 3508 TA Utrecht, The
Netherlands} 

\maketitle

\vspace*{-2pc}

\begin{abstract}
We discuss the possibility of explaining the extraordinary, correlated
X-ray/TeV flares observed during April 1997 from Mrk 501, by synchrotron-pair
cascades injected by synchrotron radiation from ultra-high energy protons and
muons. Evaluating the jet conditions required to explain the observed
features of the flares, the allowed region for this model in the parameter
space of jet magnetic field and Doppler factor is identified, and compared
with the parameter choices of other, both hadronic and leptonic models
presented in the literature. The present model requires magnetic fields
similar to other hadronic models for gamma-ray blazars ($B\gsim 30\G$), but a
lower Doppler factor ($D\approx 3$).
\end{abstract}  

\section{Correlated flares and their explanation}

In April 1997 the Beppo-SAX team observed X-ray flares from Mrk\,501
\cite{Pian98}, which are extraordinary in at least two respects: (a) they
extend to energies beyond $200\keV$, and (b) they show an extremely flat
spectrum, in one case with a power law flux index $\alpha_{\rm x}\approx
0.5$. Simultaneously, the Whipple and HEGRA Cherenkov telescopes observed
correlated flares in the TeV band \cite{TeV-501April}. Integrating over a
larger time window, both found that the high energy emission during this
period has extended on average up to at least $20\TeV$ \cite{TeVspec}, but
with a significant curvature consistent with an exponential cutoff at
${\sim}\,5\TeV$. The most common way to explain this emission is the
synchrotron-self Compton (SSC) model, which naturally expects correlated
variability because both the X-ray and the TeV component are radiated by the
same population of particles (electrons). Another appealing possibility
seems, to explain the high energy emission as synchrotron radiation from
ultra-high energy (UHE) protons, while the X-rays are produced by
co-accelerated electrons \cite{MP99ff,Aha00}.  However, both models share the
problem to explain the hard X-ray spectral index, which requires an electron
injection spectrum $dN_e/dE \propto E^{-1}$ \cite{Pian98} (this is because
synchrotron cooling steepens the stationary electron spectrum by one power
compared to the injection spectrum). However, the Fermi shock acceleration
mechanism, commonly assumed to energize the particles in the jets, cannot
produce such hard spectra (the limit is $dN/dE \propto E^{-1.5}$
\cite{Malkov}), and convincing alternatives for electron acceleration to TeV
energies in jets have not been suggested jet.

It has been pointed out previously \cite{Rac99} that correlated variability
can also find a natural explanation by considering the TeV and X-ray emission
as different generations of a synchrotron-pair cascade, injected by UHE
protons. This is a modification and extension of the proton-induced cascade
(PIC) models \cite{Man93ff}, considering the synchrotron emission of UHE
protons and muons additionally to UHE gamma-ray injection by $\pi^0$-decay.
The $p/\mu$ synchrotron component leads here to ``narrow'' cascades with an
extremely flat spectrum, which peak in the TeV and X-ray regime. It has been
shown that the asymptotic spectral indices predicted by this model for the
X-ray and TeV cascade generations fit quite well to the indices measured by
Beppo-SAX and HEGRA/Whipple, respectively \cite{Rac99}, and this for particle
injection spectra which are canonically expected from Fermi acceleration. In
contrast to pure proton-synchrotron and SSC models this model assumes that
the jet is moderately optically thick at TeV energies, to allow for a
reprocessing of the power into the X-ray regime.%
\footnote{Despite occasional, contrary claims in the literature this is not
in conflict with observations, since in a homogeneous emitter $\gamma\gamma$
absorption only induces a steepening of the power-law spectrum by the target
photon index $\alpha_{\rm t}$, rather than an exponential cutoff
\cite{Sve-MKB,Rac99}. The reader may keep this in mind when comparing the jet
Doppler factors derived here with common ``lower limits'', which all assume
that the jet emission is optically thin for all observed energies.}
In the following we shall derive the allowed region in the jet-parameter
space to explain the observed photon energies in the April 1997 flares from
Mrk 501 within this scenario, and compare with the parameter choice of other
models. A detailed fit to the flare spectra, employing the full set of
particle and photon transport equations will be presented elsewhere (Rachen
\& Mannheim, in preparation).

\section{Conditions for TeV/X-ray cascades}

\setfig{t}{par}
\epsfxsize0.75\textwidth
\centerline{\epsfbox{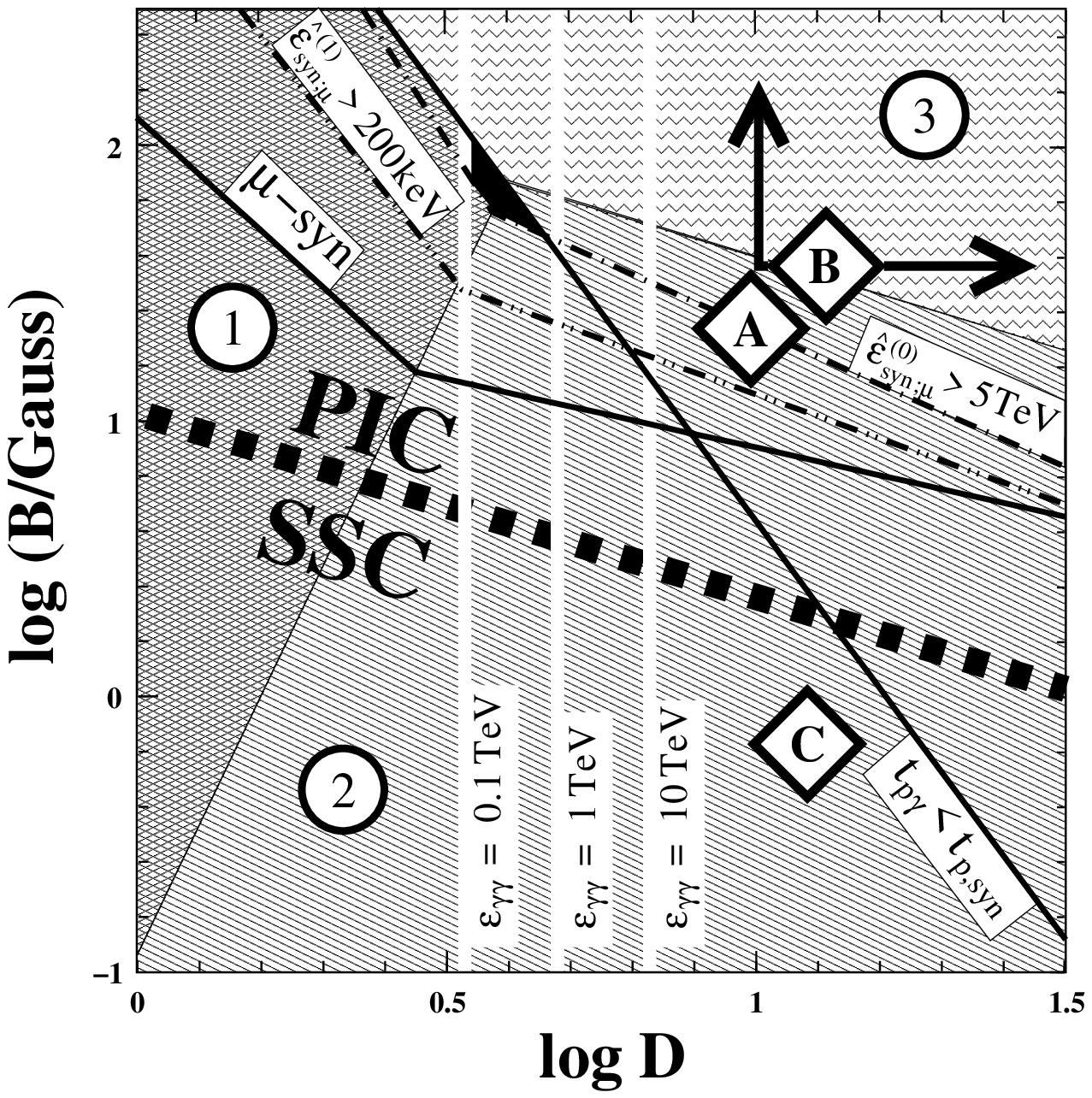}}
\figcap{\small Parameter space for the flare emission from Mrk\,501,
divided into regions of dominant processes limiting the proton energy: (1)
photohadronic losses, (2) Larmor limit, (3) synchrotron losses. Significant
muon synchrotron radiation is expected between the black solid lines labelled
$t_{p\gamma} < t_{p,\rm syn}$ and $\mu$-syn. Dot-dashed lines mark the
conditions for the maximum photon energies in the cascade, white lines the
opacity limits for cascade reprocessing (see text). Also shown are the
division line between dominant PIC and SSC emission for $L_p = L_\gamma$ (see
Ref.~\cite{Rac99}), and the jet parameters assumed in other models: proton
synchrotron models by $<$A$>$ M\"ucke \& Protheroe \cite{MP99ff} and $<$B$>$
Aharonian \cite{Aha00}, and $<$C$>$ the SSC model used in Ref.~\cite{Pian98}.}
\text
M\"ucke \& Protheroe \cite{MP99ff} have argued that the idea of reprocessing
TeV photons in a cascade to X-ray energies is incompatible with a dominant
production of TeV photons by proton-synchrotron radiation. This is because
synchrotron radiation from protons accelerated on their Larmor time scale,
which is the fastest possible for Fermi acceleration, has a high energy limit
at $\hat \eps_{{\rm syn};p} \sim 3 m_p c^2 D/\alpha_{\rm F}$, where $D$ is
the jet Doppler factor and $\alpha_{\rm F}$ the fine structure constant. The
condition $\hat \eps_{{\rm syn};p} > 5\TeV$ then requires $D>12$, which
implies for the observed luminosity from Mrk\,501 that the jet is optically
thin for TeV photons. The requirement on the jet Doppler factor can be
relaxed, however, if we consider synchrotron radiation from photohadronically
produced muons \cite{RM98}, for which $\hat\eps_{\rm syn;\mu} =
(m_p/m_\mu)\hat\eps_{{\rm syn};p}$. More precisely, the condition for
explaining the observed multi-TeV emission by muon synchrotron radiation is
\eqn{TeV}
\hat\eps^{(0)}_{\rm syn;\mu} \approx \frac{3\pi}{8} \frac{m_e^2 c^2}{m_\mu}
\frac{B \hat\gamma_p^2 D}{B_c} > 5\TeV,
\text
where $B_c = 4.4\mal 10^{13}\G$ is the critical magnetic field, $B$ is the
jet magnetic field, and $\hat\gamma_p$ is the maximum proton Lorentz factor,
which is given as a function of $B$ and $D$ depending on the dominant cooling
process (see Ref.~\cite{RM98} for details). Similarly, we can write the
condition $\hat\eps^{(1)}_{\rm syn;\mu} = (3\pi/32) (B/m_e c^2 B_c D)
[\hat\eps^{(0)}_{\rm syn;\mu}]^2 > 200\keV$ for the requirement that the next
photon generation in the synchrotron-pair cascade explains the observed X-ray
photon energies. Two more conditions arise from the assumption that muon
synchrotron radiation is relevant at all, compared to proton synchrotron
radiation: (a) the time scale for photoproduction of mesons must be at least
comparable with the synchrotron loss time of the protons ($t_{p\gamma} <
t_{p,syn}$); (b) muons can lose a significant fraction of their energy before
they decay ($t_{\mu,\rm syn} < \tau_\mu\gamma_p$). Both conditions have been
evaluated following Rachen \& M\'esz\'aros \cite{RM98}, where we assume a
spherical emission region of radius $R = [10^{15}\cm]\,D$, with an observed
(isotropic) luminosity of $10^{42}\erg\scnd^{-1}$ at $10^{12}\Hz$, and a
soft photon spectrum extending up to ${\sim}1\keV$ following $dN_{\rm
ph}/d\eps \propto \eps^{-1.85}$. This target photon spectrum is theoretically
motivated and does not match the observed infrared-to-X-ray spectrum from
Mrk\,501, which probably arises from a much larger emission volume since it
is not strongly variable, and hence does not necessarily dominate the local
photon density in the much smaller, variable emission region. At last, we
have to consider the opacity for $\gamma\gamma$ absorption, where we define
the opacity-break energy $\eps_{\gamma\gamma}$ by
\eqn{opacity}
\tau_{\gamma\gamma}(\eps_{\gamma\gamma}) = 
	\frac{0.2 \sigma_{\rm T} R m_e^2 c^4}{\eps_{\gamma\gamma}}
	\frac{dN_{\rm ph}}{dE}
		\left(\frac{m_e^2 c^4}{\eps_{\gamma\gamma}}\right) = 1\;,
\text
and require $0.1\TeV < \eps_{\gamma\gamma} < 1\TeV$, to ensure both sufficient
emission of TeV radiation and sufficient cascade reprocessing to produce the
X-ray flare.

\section{Discussion and comparison with other models}

\Fig{par} shows the conditions discussed above as limiting lines in the
parameter space of jet magnetic field $B$ and Doppler factor $D$. We see that
the condition $t_{p\gamma} < t_{p,\rm syn}$ together with Eqs.~\refeq{TeV}
and \refeq{opacity} define a small allowed region in the parameter space,
close to the ``star-point'' of equal cooling times defined by Rachen \&
M\'esz\'aros \cite{RM98}. Other conditions turn out to be less restrictive
for the present case. The result of M\"ucke \& Protheroe \cite{MP99ff}, who
also considered muon-synchrotron radiation and cascade reprocessing in their
MC simulations and found both effects insignificant, can thus be understood
by their parameter choice (marked as point $<$A$>$ in \fig{par}), which
implied $t_{p,\rm syn}\ll t_{p\gamma}$ and $\eps_{\gamma\gamma}>10\TeV$. We
note that the jet Doppler factor determined here ($D\approx 3$) is
signficantly lower than those assumed in both pure proton-synchrotron and SSC
models.
    
The small allowed parameter region would also explain why flares like those
seen in April 1997 are rather rare events. We emphasize that explaining TeV
radiation alone by hadronic phenomena, like PIC or proton-synchrotron
radiation, is possible for a much larger set of parameters. Also correlated
variability may arise from less restrictive settings, for example by
correlated acceleration of protons and electrons in the jet \cite{MP99ff}.
Finally, the fact that the different models are so clearly separated in the
parameter space opens the chance to distinguish between them by determining
the jet Doppler factor and magnetic field {\em independent} of the assumed
radiative mechanism.

\begin{small}

\vspace{8pt}

\noindent {\bf Acknowledgments.} The author is indebted to Karl Mannheim for
his contributions to this research, and to the EU-TMR network
``Astro-Plasmaphysics'' (ERBFMRX-CT98-0168) for support.

\end{small}


\vspace{-18pt}

\begin{references}

\bibitem{Pian98}
E. Pian {\it et~al.}, {\apjl} {\bf 497}, L17  (1998).

\bibitem{TeV-501April} M.~Catanese {\it et al.}, {\apj} {\bf 487}, 143 (1997),
F.A.~Aharonian {\it et al.}, {\aap} {\bf 342}, 69 (1999).

\bibitem{TeVspec} F. Krennrich {\it et al.}, {\apj} {\bf 511}, 149 (1998);
A. Konopelko, {\apph} {\bf 11}, 135 (1998).

\bibitem{MP99ff} A. M\"ucke and R.J. Protheroe, in {\it GeV-TeV Gamma Ray
Astrophysics Workshop, Snowbird, Utah, 1999}, edited by B.L.~Dingus {\it et
al.}, AIP Conf.~Proc.~ No.~515 (2000), pp. 149--153, astro-ph/9910460;
{\apph}, in press, astro-ph/0004052.

\bibitem{Aha00} F.A. Aharonian, astro-ph/0003159.

\bibitem{Malkov} M.A.~Malkov, {\apjl} {\bf 511}, L53 (1999).

\bibitem{Rac99} J.P. Rachen, Snowbird Proceedings (see \cite{MP99ff}),
pp.~41--52, astro-ph/0003282.

\bibitem{Man93ff} K.\,{Mannheim}, {\aap}\,{\bf 269},\,67\,(1993);
Sp.\,Sci.\,Rev.\,{\bf 75},\,331\,(1996); {\science}\,{\bf
279},\,684\,(1998).

\bibitem{Sve-MKB} R.~Svensson, {\mnras} {\bf 227}, 403 (1987); K. {Mannheim}
{\it et al.}, {\aap} {\bf 251}, 723 (1991).

\bibitem{RM98} J.P. {Rachen} and P. {M{\'e}sz{\'a}ros}, {\physrevd} {\bf
58}, 123005 (1998).

\end{references}
\end{document}